\documentclass[useAMS,usenatbib]{mn2e}
\usepackage{times}
\usepackage{graphics,epsfig}
\usepackage{graphicx}
\usepackage{amssymb}

\newcommand{\vU}{{\vec{ U}}}
\def\V{{\cal V}}

\title[H~{\sc i} opacity fluctuations power spectra]{H~{\sc i} 21~cm opacity fluctuations power spectra towards Cassiopeia A} 
\author[N. Roy et al.]{Nirupam Roy $^{1}$\thanks{E-mail: nroy@aoc.nrao.edu~(NR); chengalu@ncra.tifr.res.in~(JNC); prasun@cts.iitkgp.ernet.in~(PD); somnath@cts.iitkgp.ernet.in~(SB)}, Jayaram N. Chengalur $^{2}$\footnotemark[1], Prasun Dutta $^{3}$\footnotemark[1] and Somnath Bharadwaj $^{3}$\footnotemark[1]\\ 
       $^{1}$National Radio Astronomy Observatory, 1003 Lopezville Road, Socorro, NM 87801, USA\\
       $^{2}$National Centre for Radio Astrophysics, TIFR, Post Bag 3, Ganeshkhind, Pune 411 007, India\\
       $^{3}$Department of Physics and Meteorology \& Centre for Theoretical Studies, IIT Kharagpur, Kharagpur 721 302, India}

\begin{document}
\date{Accepted yyyy month dd. Received yyyy month dd; in original form yyyy 
month dd}

\pagerange{\pageref{firstpage}--\pageref{lastpage}} \pubyear{2010}

\maketitle

\label{firstpage}

\begin{abstract}
The angular power spectrum of H~{\sc i} 21~cm opacity fluctuations is a useful 
statistic for quantifying the observed opacity fluctuations as well as for 
comparing these with theoretical models. We present here the H~{\sc i} 21~cm 
opacity fluctuation power spectrum towards the supernova remnant Cas~A from 
interferometric data with spacial resolution of $5\arcsec$ and spectral 
resolution of $0.4$~km~s$^{-1}$. The power spectrum has been estimated using a 
simple but robust visibility based technique. We find that the power spectrum 
is well fit by a power law $P_{\tau}(U) = U^{\alpha}$ with a power law index 
of $\alpha \sim -2.86 \pm 0.10$ ($3\sigma$ error) over the scales of $0.07 - 
2.3$ pc for the gas in the Perseus spiral arm and $0.002 - 0.07$ pc ($480 - 
15730$ au) for that in the Local arm. This estimated power law index is 
consistent with earlier observational results based on both H~{\sc i} emission 
over larger scales and absorption studies over a similar range of scales. We 
do not detect any statistically significant change in the power law index with 
the velocity width of the frequency channels. This constrains the power law 
index of the velocity structure function to be $\beta = 0.2 \pm 0.6$ ($3\sigma$ 
error).
\end{abstract}


\section{Introduction}
\label{sec:hip-int}

It is now well established that the H~{\sc i} 21~cm opacity in the Galaxy 
shows measurable small scale structure. Very Long Baseline Interferometry 
(VLBI) observations show opacity fluctuations on angular scales as small as 20 
milli arc seconds corresponding to a projected separation $\sim 10$ au, if one 
assumes that the opacity variations occur in a thin screen \citep{br05}. 
Similarly, multi-epoch observations of some high velocity pulsars show 
evidence for H~{\sc i} 21~cm opacity variation on scales of 5 - 100 au 
\citep{fr94}. Several other pulsars however do not show opacity variations 
\citep{jo03}, leading to suggestions that the occurrence of fine scale 
structure in the Galaxy may be rare. Much of these observed fluctuations on au 
scales were earlier believed to originate in H~{\sc i} ``clouds'' with 
densities $\sim 10^4 - 10^5$ cm$^{-3}$. It is hard to explain the existence of 
such structures in equilibrium with other, orders of magnitude lower density, 
components of the diffuse interstellar medium (ISM). \citet{d00} has shown 
that the observed small scale transverse variations have contributions from 
structure on all scales and that the fluctuations at larger spatial scales 
could, in projection, produce the measured small scale fluctuations. 
Conversely, there are other observational and numerical simulation results 
supporting the existence of tiny scale structures which have a size of $\sim 
3000$ au, with a density of $\sim 100$ cm$^{-3}$ 
\citep[e.g.][]{bk05,va06,ha07}. With an evaporation timescale of $\sim 1$ Myr, 
these structures can survive only if the ambient pressure is much higher than 
the normal value or if they are formed continuously on a comparable timescale. 
Details regarding their origin and physical properties are, however, still not 
understood. On somewhat larger scales, \citet{de00} measured opacity 
variations of the H~{\sc i} absorption across Cassiopeia A and Cygnus A, and 
found that the power spectrum of these fluctuations could be fairly well 
represented by a power law. The slope of the power law was similar for the 
absorption in the Perseus arm (towards Cas~A) and the Outer arm (towards 
Cygnus A), but substantially different for that from absorption towards Cygnus 
A arising in the Local arm. 

The scale free behavior of fluctuation power spectra of a variety of tracers 
(H~{\sc i} 21~cm emission intensity, dust emission) is the main observational 
evidence for the existence of turbulence in the atomic ISM. The slope of the 
power law of H~{\sc i} 21~cm emission and absorption in our own Galaxy, 
H~{\sc i} 21~cm emission from the Large Magellanic Cloud, the Small Magellanic 
Cloud and DDO~210 are all $\sim -3$ \citep{cd83,gr93,st99,el01,ba06}. Recently 
\citet{dp09a,dp09b} have reported that the H~{\sc i} 21~cm intensity 
fluctuation power spectra for a sample of dwarf and nearby spiral galaxies 
have a power law index of $\sim -2.6$ and $\sim -1.5$ on scales respectively 
smaller and larger than the scale height of the galaxy disk. This is 
interpreted as the effect of a transition from three dimensional to two 
dimensional turbulence at scales larger than the disk thickness. The observed 
fractal structure of H~{\sc i} in several dwarf galaxies in the M~81 group is 
also consistent with the self-similar hierarchical structure of the turbulent 
ISM without any preferred length scale \citep{we99}.

On the theoretical side, fine scale structure in the neutral gas is naturally 
expected in turbulent models of the ISM. The fluctuations in the H~{\sc i} 21 
cm opacity in a particular velocity range depend on the fluctuations in the 
density, spin temperature and velocity of the gas. \citet{de00} show that in 
the case of small fluctuations of density, the dependence on temperature is 
small. Turbulence however, gives rise to fluctuations in both the density and 
velocity of the gas and both of these contribute to the observed opacity 
fluctuations. \citet{lp00} show that the slope of the observed power spectrum 
changes depending on whether the H~{\sc i} 21~cm emission is averaged over a 
velocity range that is large (``thick slices'') or small (``thin slices'') 
compared to the turbulent velocity dispersion. Observations with high enough 
spectral resolution can hence disentangle the statistical properties of the 
velocity and the density fluctuations. 

Cassiopeia A or Cas~A (G111.7$-$2.1) is in the constellation Cassiopeia in the 
Galactic second quadrant. It is a shell type supernova remnant of diameter 
$5\arcmin$ at a distance of approximately $3.4$ kpc \citep{re95}, and is one 
of the brightest radio sources in the sky (flux density of $\sim 2720$ Jy at 
$1$ GHz). The radio images show it to have a clear shell like structure along 
with compact emission knots, the shell thickness estimated from the radial 
brightness profile is $\sim 30\arcsec$. There have been many studies of the 
ISM towards Cas~A using H~{\sc i} 21~cm absorption 
\citep[e.g.][]{crw62,cl65,sc86,bi91,rey97} and a variety of other tracers like 
H$_2$CO and OH \citep{go84,bi86} all of which reveal small scale structure. A 
recently developed formalism for statistically robust extraction of the 
fluctuation power spectrum from interferometric data is used here to 
re-examine the issue and to constrain the H~{\sc i} 21~cm opacity fluctuation 
power spectrum towards Cas~A. This method is particularly useful because one 
can completely avoid the complications of deconvolution and imaging the strong 
background source. We note that the opacity power spectrum towards Cas~A has 
been obtained earlier by \citet{de00}. They used the data from the Very Large 
Array (VLA) B, C and D configurations but with only 18 antennas in each 
configuration due to software restrictions. Estimating the power spectrum 
using the visibility based formalism for the same line of sight will allow us 
to compare the results from two very different methods to cross-check 
different techniques. Section \S\ref{sec:data} briefly describes the data 
while our methods of estimating the intensity and opacity fluctuation power 
spectra are outlined in Section \S\ref{sec:hiatech} and \S\ref{sec:hioppk} 
respectively. Section \S\ref{sec:result} contains the results and discussion, 
and we present conclusions in Section \S\ref{sec:hicon}.

\section{Data}
\label{sec:data}
\begin{figure}
\begin{center}
\includegraphics[scale=0.65, angle=0.0]{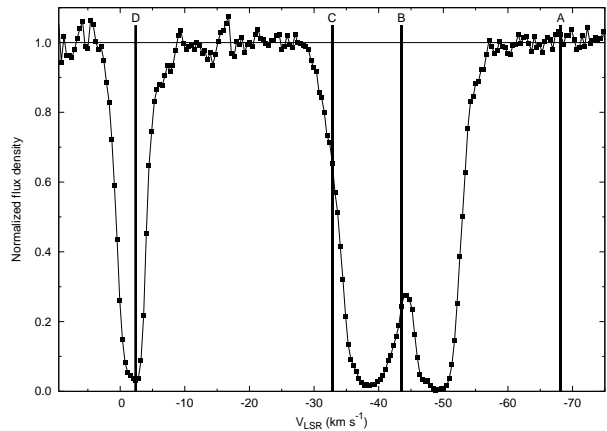}
\caption{\label{fig:hips-1} Integrated H~{\sc i} 21~cm absorption spectrum towards the supernova remnant Cas~A. The continuum corresponds to the total flux density of Cas~A. The wide feature at negative LSR velocity is the absorption produced by gas from the Perseus spiral arm and the absorption near zero LSR velocity is from the Local arm gas. Power spectra for channels marked as A, B, C and D 
are shown in Figure \ref{fig:pk}.}
\end{center}
\end{figure}
The Giant Metrewave Radio Telescope (GMRT) L-band ($21$ cm) receiver was used 
to observe the H~{\sc i} in absorption towards supernova remnant Cas~A. The 
observation was carried out on December 03, 2006 and the total duration was 
about 13 hours with on-source time of about 8 hours. VLA calibrator sources 
2148$+$611 and 2355$+$498 were used for phase calibration. One of the phase 
calibrators was observed for 3 minutes for every 20 minutes observation of 
Cas~A. Standard flux calibrators 3C48 and 3C286 were observed for about 20 
minutes in every 4 hours during the observation. Frequency switching was used 
to calibrate the spectral baseline --- the flux calibrator scans (which were 
also used for bandpass calibration) were observed at frequency offset by 5~MHz 
from that of the absorption line. A total baseband bandwidth of $0.5$~MHz 
divided into $256$ frequency channels centered at $1420.5973$~MHz was used for 
the observation. This corresponds to a velocity resolution of $\sim 
0.4$~km~s$^{-1}$ per channel and the total bandwidth was sufficient to cover 
the H~{\sc i} 21~cm absorption produced by gas both from the Perseus arm and 
the Local arm. Data analysis was carried out using standard {\small AIPS}. 
After flagging out bad data, the flux density scale, instrumental phase and 
frequency response were calibrated. The calibrated visibility data of Cas~A 
were then used to estimate the angular power spectra and the errors. The 
integrated H~{\sc i} 21~cm absorption spectrum towards Cas~A derived from the 
GMRT observation is shown in Figure (\ref{fig:hips-1}). The wide feature at 
negative velocity with respect to the Local Standard of Rest (LSR) is the 
absorption produced by gas from the Perseus spiral arm and the absorption near 
zero LSR velocity is from the Local arm gas. 

\section{The intensity fluctuation power spectrum}
\label{sec:hiatech}
The angular power spectrum $P_S(U)$ of the sky brightness fluctuations 
$\delta S(l,m)$, is the Fourier transform of the autocorrelation function 
$\xi_S$ 
\begin{equation}
P_S(U) \equiv P_S(u,v)=\int\int\xi_S(l, m)e^{-2\pi i(ul+vm)}dldm \,.
\end{equation}
Here $(l,m)$ refer to the directions on the sky, $(u,v)\equiv \vU$ refer to 
the inverse angular separation and $\xi_S$ is given by 
\begin{equation}
\xi_S(l-l^{\prime}, m-m^{\prime})=\langle\delta S(l,m)\delta S(l^{\prime},
m^{\prime})\rangle \,.
\end{equation}
The angular bracket refers to an ensemble average over different realisations 
of $\delta S(l,m)$ which is assumed to be stochastic in nature. The 
fluctuations are also assumed to be statistically isotropic which implies that 
$P_S(U)$ depends only on $U=\mid \vU \mid$. 

The visibility $\V_{\nu}(\vU)$ measured in radio interferometric observations 
directly probes a Fourier mode of the sky brightness fluctuation $\delta 
S(l,m)$. Here the inverse angular scale $\vU$ directly corresponds to a 
baseline, the antenna separation projected in the plane perpendicular to the 
direction of observation, measured in units of the observing wavelength 
$\lambda$. It is common practice to express the dimensionless quantity $\vU$ 
in units of kilo~wavelength (k$\lambda$). The Fourier relation between 
$\V_{\nu}(\vU)$ and $\delta S(l,m)$ allows us to identify each baseline with 
an inverse angular scale, and to directly estimate the angular power spectrum 
from the measured visibilities. 

Detailed discussions of the visibility based power spectral estimator that we 
use here can be found in \citet{ba06} and \citet{dp09a}. Briefly, the 
estimator $\hat{\rm P}(\vU)=\langle\ \V_{\nu}(\vU)\V^{*}_{\nu}(\vU+\Delta 
\vU)\ \rangle$, is obtained by correlating every visibility $\V_{\nu}(\vU)$ 
with all other nearby visibilities $\V^{*}_{\nu}(\vU+\Delta \vU)$, i.e. those 
which lie within the disk defined by $|\Delta \vU| < (\pi \theta_{0})^{-1}$, 
where $\theta_0$ is the angular radius of the source. In the case of Cas~A, 
$\theta_0 \sim 2.5'$, i.e. the angular radius of the source. Since the 
fluctuations are assumed to be statistically isotropic, the correlations are 
averaged over different $\vU$ directions to obtain the final estimator. To 
increase the signal to noise ratio we further average the estimator in bins of 
$U$. \citet{dp09a} show that the expectation value of the estimator 
$\hat{\rm P}(\vU)$ is the convolution of the power spectrum $P(U)$ with a 
window function $|\tilde{W}_{\nu}(U)|^{2}$. The window function, which 
quantifies the effect of the finite angular extent of the source, is peaked 
around $U=0$ and has a width of order $(\pi\theta_{0})^{-1}$ beyond which 
$|\tilde{W}_{\nu}(U)|^{2}\sim 0$. For the optically thin shell-type geometry 
of Cas~A the convolution does not effect the shape of the power spectrum 
beyond a baseline $U_m\approx1.6$~k$\lambda$ \citep{nr09}, and we may directly 
interpret the real part of the estimator $\hat{\rm P}(\vU)$ as the power 
spectrum $P(U)$. The estimator $\hat{\rm P}(\vU)$ also has a small imaginary 
component arising mainly from noise. We compute the $1\sigma$ error bars for 
the estimated power spectrum by assuming the error to be the quadrature sum of 
contributions from two sources of uncertainty, viz. the sample variance and 
the system noise. At small $U$ the uncertainty is dominated by the sample 
variance which comes from the fact that we have a finite and limited number of 
independent estimates of the true power spectrum. At large $U$, it is 
dominated by the system noise in each visibility. 

The longest baseline in our observation ($\sim 25$ km) corresponds to an 
angular resolution of $\sim 2.5\arcsec$, but our power spectrum estimation is 
restricted to $U \le 50$~k$\lambda$ (angular scale $\sim 5\arcsec$) because 
the larger baselines do not have an adequate signal to noise ratio. Similarly 
for $U \le 1.6$~k$\lambda$ (angular scale $\sim 2.5\arcmin$) the sample 
variance leads to large uncertainties in the power spectral estimate. Note 
that since the estimator is local in $U$, the estimate of the power 
spectrum over this specific range in $U$ is not affected by the missing zero 
spacing baselines or the cut-off at large $U$. Certainly these cut-offs and 
consequent Gibbs phenomena will produce significant ringing and distortion in 
the image. However, since the power spectrum is estimated directly in the $uv$ 
plane, our method, unlike image based techniques, has completely avoided these 
problems. 

\begin{figure}
\begin{center}
\includegraphics[scale=0.65, angle=0.0]{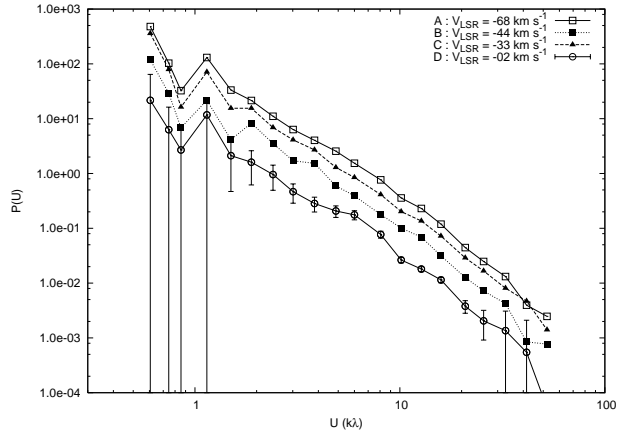}
\caption{\label{fig:pk} Power spectra for channels marked as A, B, C and D in Figure \ref{fig:hips-1}. Typical error bars are plotted for only one channel for clarity. Since channel A has no H~{\sc i} absorption, it just shows intensity fluctuations power spectrum of the background source Cas~A. Channel D has inadequate signal to noise to constrain the power spectrum due to very large optical depth. Model fit to the data of channel C (V$_{\rm LSR} = -33$~km~s$^{-1}$) is shown in Figure \ref{fig:hips-2}.}
\end{center}
\end{figure}

Figure \ref{fig:pk} shows the estimated power spectrum for four different 
channels with H~{\sc i} absorption features of different optical depth. Note 
that since channel A has no H~{\sc i} absorption, it just shows intensity 
fluctuations power spectrum of the background source Cas~A. On the other hand, 
channel D has inadequate signal to noise to constrain the power spectrum due 
to very large optical depth. The power spectrum $P_{I_c}$ of the continuum 
emission from Cas~A, seen in the channels without H~{\sc i} absorption, has 
been studied in \citet{nr09}. This power spectrum is well fit by a broken 
power law 
\begin{eqnarray}
P_{I_c} &=& C U^{-2.22} \quad {\rm for} \quad U < 10.6 \quad {\rm
k}\lambda \nonumber \\ 
 &=& 10.6^{1.01} C U^{-3.23} \quad {\rm for} \quad U > 10.6
\quad {\rm k}\lambda \label{eq:model} \,
\end{eqnarray} 
which has been interpreted as arising from magnetohydrodynamic turbulence in 
this shell type supernova remnant. The angular scale of the break in the power 
spectrum corresponds to the shell thickness of Cas~A \citep{nr09}.

\section{The opacity fluctuation power spectrum}
\label{sec:hioppk}

The sky brightness $S(l,m)$ towards the extended source Cas~A may be written as 
\begin{equation}
S(l,m)=I_c(l,m) e^{-\tau(l,m)}
\end{equation}
where $I_c(l,m)$ is the continuum radiation from Cas~A and $\tau(l,m)$ is the 
optical depth for absorption by intervening H~{\sc i}. We expect $I_c(l,m)$ 
and $\tau(l,m)$ to be independent, whereby we can write
\begin{equation}
\xi_S=\langle I_c(l,m)I_c(l^{\prime}, m^{\prime})\rangle \times \langle e^{-\tau(l,m)}e^{-\tau(l^{\prime}, m^{\prime})}\rangle.
\label{eqn:interm1}
\end{equation}
Both $I_c(l,m)$ and $\tau(l,m)$ can be decomposed into the sum of a constant 
average value and the fluctuation around that value (as a function of $l,m$) 
\begin{eqnarray}
I_c(l,m)&=&I_c^0 + \delta I_c(l,m) \nonumber \\
\tau(l,m)&=&\tau^0 + \delta \tau(l,m)
\end{eqnarray}
and hence equation (\ref{eqn:interm1}) can be re-written as
\begin{equation}
\xi_S=(I_c^{0\,2}+\xi_{I_c})\times e^{-2\tau^0}\langle e^{-\delta\tau(l,m)}e^{-\delta\tau(l^{\prime}, m^{\prime})}\rangle
\end{equation}
where $\xi_{I_c}$ is the autocorrelation function of continuum intensity 
fluctuation. Further, we assume that the opacity fluctuation $\delta\tau$ is a 
Gaussian random field with zero mean, variance $\sigma^2_{\tau}$ and 
autocorrelation function $\xi_{\tau}$, whereby
\begin{equation}
\langle e^{-\delta\tau(l,m)} e^{-\delta\tau(l^\prime,m^\prime)}
\rangle = \exp(\sigma_{\tau}^2+ \xi_{\tau}) \,. 
\end{equation}
The opacity power spectrum $P_{\tau}={\mathcal F}(\xi_{\tau})$, which is our 
main interest in this paper, is the Fourier transform of $\xi_{\tau}$. The 
power spectrum of the brightness fluctuations for a frequency channel with 
H~{\sc i} absorption can be written as
\begin{equation}
P_S={\mathcal
 F}\{\xi_S\}=e^{-2\tau^0+\sigma_\tau^2}(I_c^{0\,2}+P_{I_c})\otimes{\mathcal
 F}\{e^{\xi_{\tau}}\} \,
\label{eq:psf}
\end{equation}
where $P_{I_c}$ is the power spectrum of brightness fluctuations of the 
continuum emission, i.e. the Fourier Transform of $\xi_{I_c}$. Equation 
(\ref{eq:psf}) relates $P_S$ to the convolution of $(I_c^{0\,2}+P_{I_c})$ and 
$\mathcal{F}\{e^{\xi_{\tau}} \}$ which refer to the continuum and the 
H~{\sc i} opacity respectively. It should be noted that while the terms $P_S$, 
$P_{I_c}$ and $\mathcal{F}\{e^{\xi_{\tau}} \}$ are all functions of $\vU$, the 
terms $I_c^{0}$ and $e^{-2\tau^0+\sigma_\tau^2}$ do not have any $\vU$ 
dependence. In our analysis we have treated $B = e^{-2\tau^0+\sigma_\tau^2}$ 
as a free parameter that determines the amplitude of $P_S$. The terms 
$I_c^{0\,2}$ and $P_{I_c}$, that refer to the continuum radiation, are known 
from Cas~A's total continuum flux density and analysis of the brightness 
fluctuations in the line free channels (i.e. eq. \ref{eq:model}) respectively. 
The H~{\sc i} opacity power spectrum $P_{\tau}$, which is the quantity that we 
wish to determine in this analysis, enters eq. (\ref{eq:psf}) through the term 
${\mathcal F}\{e^{\xi_{\tau}}\}$. We have assumed that the opacity power 
spectrum is a power law
\begin{equation}
P_{\tau} = {\mathcal F}\{\xi_{\tau}\} = AU^{\alpha}\,.
\label{eq:ptau}
\end{equation}
Under this assumption the brightness power spectrum $P_S(U)$ is completely 
determined by three parameters $A, B$ and $\alpha$. For each velocity channel 
that shows H~{\sc i} absorption we have used standard chi-square minimization 
to determine the value of these three parameters for which our model 
prediction (eqs. \ref{eq:psf} and \ref{eq:ptau}) best fits the measured $P_S$. 
The values of $A$ and $B$ are not of particular interest in the present 
analysis. Hence we derive only the best fit value of $\alpha$ and marginalize 
over $A$ and $B$ to estimate errors for $\alpha$. 

\section{Results and Discussion}
\label{sec:result}
\begin{figure}
\begin{center}
\includegraphics[scale=0.65, angle=0.0]{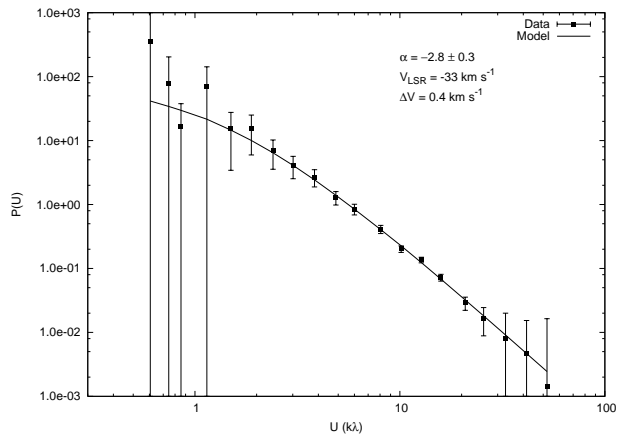}
\caption{\label{fig:hips-2} Intensity fluctuation power spectrum for a ``thin'' velocity channel with H~{\sc i} absorption (channel C marked in Figure \ref{fig:hips-1} and Figure \ref{fig:pk}). The best fit model power spectrum is plotted with the data points.}
\end{center}
\end{figure}

Figure \ref{fig:hips-2} shows the measured $P_S$ and the best fit model of 
$P_S$ for a single channel with H~{\sc i} absorption. We find that our model, 
which assumes a power law H~{\sc i} opacity power spectrum (eq. \ref{eq:ptau}), 
provides a good fit to the measured sky brightness power spectrum $P_S$. The 
reduced chi-square values for the best fit model are, however, not close to 
$1.0$ but are about $0.3$. This is most probably because of an overestimation 
of the errors in the measured $P_S$. The estimated errors, for example, are 
found to be significantly more than the variation of the continuum power 
spectra across line free channels. It implies that the very conservative and 
approximate error estimates adopted here may result in a smaller value of the 
reduced chi-square even when the model represents the observed power spectrum 
quite well. Hence, we have scaled down the errors of the observed $P_S$ so 
that the reduced chi-square is close to $1.0$. The $1\sigma$ errors of the 
power law index correspond to a change of the reduced chi-square by $+1$ from 
its minimum value.

The velocity channels with very strong absorption do not have adequate signal 
to noise to constrain the power spectrum. On the other hand, for a few 
channels with dominant continuum signal and very weak H~{\sc i} absorption, 
the power spectrum does not deviate significantly from the ``continuum'' power 
spectrum, making it hard to constrain the opacity spectrum. The optimal 
channels for this analysis are hence those with moderate H~{\sc i} absorption, 
where there is still sufficient signal to noise ratio, and the total power 
spectrum has been significantly modified by the H~{\sc i} absorption. We note 
that any image-based method of estimating the power spectrum will also be 
constrained to the same optimal channels by the signal to noise considerations 
mentioned here. 

The best fit value of $\alpha$ is found to be $-2.86 \pm 0.30$ ($3\sigma$ 
error) for the angular scale of $5 - 150\arcsec$ probed in this observation. 
Note that this error bar is based on the quadrature sum of the estimated noise 
due to ``cosmic variance'' and the signal to noise ratio on each visibility. 
As discussed above, this is likely to be an overestimate of the true error. 
Consistent with this, the channel to channel variation of the best fit value 
of $\alpha$ is much less than the estimated error of $0.30$. Computing the 
variance over independent estimates of $\alpha$ derived for different velocity 
channels leads to $\alpha \approx -2.86 \pm 0.10$ ($3\sigma$ error). We find 
no significant difference in $\alpha$ between the velocity channels with 
absorption produced by gas from the Perseus arm and from the Local arm. The 
angular scales that our power spectral estimator is valid for corresponds to a 
linear scale of $0.07 - 2.3$~pc for the Perseus arm (for an assumed distance 
of 3 kpc to the arm). The Local arm is $\sim 100$ pc wide in the solar 
neighbourhood \citep{pr98}. So, for the Local arm, the upper limit of linear 
scale range (adopting a distance of $100$ pc to the absorbing gas) is $0.002 - 
0.07$ pc ($480 - 15730$ au). We note that, for the Local arm towards Cygnus~A, 
\citet{de00} have reported $\alpha = -2.5$, an index somewhat shallower than 
what we find here, or what those authors find for the Perseus arm and the 
Outer arm. For the Perseus arm towards Cas~A, \citet{de00} report $\alpha = 
-2.75 \pm 0.25$ ($3\sigma$ error) for a similar range of linear scales. This 
factor of $\sim 2.5$ smaller uncertainty in the current measurement is partly 
due to the better sensitivity of the present observations and partly due to 
using a more optimal estimator of the power spectrum. The value of $\alpha$ 
that we find here is also consistent with the value ($\sim -3$) reported from 
the Galactic H~{\sc i} emission observations \citep{gr93} probing the scales 
of $50 - 200$~pc. 

For power law opacity fluctuations power spectrum (eq. \ref{eq:ptau}) with 
$-2 \ge \alpha \ge -4$, the opacity fluctuation smoothed over a length-scale 
$x$ is predicted \citep{d00} to have a rms $\sigma_{\tau}(x) \propto 
x^{\frac{\alpha-2}{2}} $ which is also a power law. Using our best fit model 
we have $\sigma_{\tau}(x) = 2.0~\, (x/4~\, {\rm pc})^{0.43}$. Extrapolating 
this power law, the rms opacity variation is $\sim 0.05$ on scales of $100$ au 
which is broadly consistent with recent VLA and VLBA studies \citep{fg98,de00,fg01,br05,la09}. The present study of H~{\sc i} absorption, combined 
with this result derived from H~{\sc i} emission observations at larger linear 
scales indicates that there is no significant change in the slope of the power 
spectrum over $4-5$ orders of magnitude in linear scale. This spectrum is 
significantly shallower than the Kolmogorov spectrum (with power law index of 
$11/3$) expected from an incompressible turbulent medium \citep{ko41}. This 
may be because of the fact that the turbulence in H~{\sc i} is compressible 
and magnetohydrodynamic in nature \citep[see][for discussion on this 
issue]{nr09}. 

If the density fluctuation is small compared to the mean density, then for gas 
in pressure equilibrium, the slope of the opacity fluctuation power spectrum 
will be nearly the same as that of the density fluctuation power spectrum 
\citep{de00}. However, fluctuations in both the density and velocity fields 
contribute to the observed H~{\sc i} opacity fluctuations. It can be shown 
that for the power spectrum estimated from ``thick slices'', i.e. those with 
velocity width larger than the turbulent velocity dispersion, the contribution 
is only from the density fluctuations, and all velocity information get 
averaged out \citep{lp00}. So, for ``thick slices'', the intensity fluctuation 
is dominantly due to density fluctuation and the observed power law index $n$ 
is expected to be same as the index of the density fluctuation power spectrum. 
On the other hand, for ``thin slices'', the power law index of the observed 
power spectrum is $n + \beta/2$, where $\beta$ is the power law index of the 
velocity structure function. Hence, it is possible to decouple the density and 
velocity fluctuations from the power spectra derived from ``thin'' and 
``thick'' velocity channels. \citet{lp00} reported a near-Kolmogorov power law 
index for both density and velocity fluctuations power spectra for the Milky 
Way and the Small Magellanic Cloud by applying this technique to the H~{\sc i} 
emission data \citep{gr93,st99}. For our data, we find that though there is a 
very weak trend of smaller $\alpha$ for larger channel width, the observed 
change of $\alpha$ by about $-0.1$ for a change of velocity width from $\sim 
0.4$ to $\sim 12.8$~km~s$^{-1}$ is within the $1~\sigma$ error bar. Since 
$\alpha$ remains unchanged for ``thin slices'' with velocity width as small as 
$\sim 0.4$ km~s$^{-1}$, i.e.~much smaller than the typical value of 
$4.0$~km~s$^{-1}$ for turbulent dispersion in Galactic cold H~{\sc i} 
\citep{ra72}, we can constrain $\beta$ to be $0.2 \pm 0.6$ ($3\sigma$ error). 
This is consistent with the value of $\beta = 2/3$ predicted for turbulence in 
an incompressible medium \citep{ko41}. 

\section{Conclusions}
\label{sec:hicon}

In this work, we have studied the H~{\sc i} 21~cm opacity fluctuation towards 
Cas~A. A simple but robust method of directly estimating the power spectrum 
from the observed visibilities is outlined. In this analysis we avoid the 
complications of imaging the bright, extended continuum source, of subtracting 
the continuum and of making optical depth image for channels with H~{\sc i} 
absorption.

We have found that the H~{\sc i} opacity fluctuation power spectrum can be 
modelled as a power law with an index of $-2.86 \pm 0.10$ ($3\sigma$ error). 
This is consistent with earlier observational results. We have not found any 
significant difference of the power law index between velocity channels with 
absorption produced by the gas from the Perseus arm and the Local arm. We have 
also checked, by smoothing the visibility data for adjacent channels, if there 
is variation of the power law index with the velocity width of channels. It is 
found that, within the estimation errors, the power law index remains constant 
for a wide range of velocity widths. We can not, however, rule out 
contribution of velocity fluctuations to the observed opacity fluctuations if 
the power law index of the velocity structure function is $0.2 \pm 0.6$ 
($3\sigma$ error). We plan to apply, in future, this visibility based method 
to study the opacity fluctuations for other lines of sight in different 
regions of the Galaxy.

\section*{Acknowledgements}

We thank the staff of the GMRT who have made these observations possible. 
%NR 
%would like to acknowledge 
%the hospitality of all the staff members of the 
%Centre for Theoretical Studies (IIT, Kharagpur) during his stay for this 
%collaboration, the support during his stay at NCRA-TIFR where a major fraction 
%of this research was carried out, and 
%the support of the National Radio 
%Astronomy Observatory (NRAO). The NRAO is a facility of the National Science 
%Foundation operated under cooperative agreement by Associated Universities, 
%Inc. 
PD would like to acknowledge HRDG CSIR and SRIC, IIT, Kharagpur for 
providing financial support. S.B. would like to acknowledge financial support 
from BRNS, DAE through the project 2007/37/11/BRNS/357. We are grateful to 
Rajaram Nityananda and Sanjit Mitra for many helpful comments. We are also 
grateful to the anonymous referee for prompting us into substantially 
improving this paper.

%\bsp

\label{lastpage}

\end{document}